# A Stimulated Raman Loss spectrometer for metrological studies of quadrupole lines of hydrogen isotopologues.

M. Lamperti[1,*], L. Rutkowski[2], D. Gatti[1], R. Gotti[1,†], L. Moretti[1], D. Polli[1], G. Cerullo[1], M. Marangoni[1]

[1] *Dipartimento di Fisica - Politecnico di Milano and IFN-CNR, Via Gaetano Previati 1/C, 23900 Lecco, Italy*

[2] *Univ Rennes, CNRS, IPR (Institut de Physique de Rennes)-UMR 6251, F-35000 Rennes, France*

\* *marco.lamperti@uninsubria.it, currently at Dipartimento di Scienza e Alta Tecnologia – Università degli studi dell'Insubria, via Valleggio 11, 22100 Como, Italy*

[†]*Currently at Dipartimento di Ingegneria Industriale e dell'Informazione, Università di Pavia, Via Ferrata 5, 27100 Pavia, Italy*

**Abstract**

We discuss layout and performance of a high-resolution Stimulated Raman Loss spectrometer that has been newly developed for accurate studies of spectral lineshapes and line center frequencies of hydrogen isotopologues and in general of Raman active transitions. Thanks to the frequency comb calibration of the detuning between pump and Stokes lasers and to an active alignment of the two beams, the frequency accuracy is well below 100 kHz. Over the vertical axis the spectrometer benefits from shot-noise limited detection, signal enhancement via multipass cell, active flattening of the spectral baseline and measurement times of few seconds over spectral spans larger than 10 GHz. Under these conditions an efficient averaging of Raman spectra is possible over long measurement times with minimal distortion of spectral lineshapes. By changing the pump laser, transitions can be covered in a very broad frequency span, from 50 to 5000 cm$^{-1}$, including both vibrational and rotational bands. The spectrometer has been developed for studies of fundamental and collisional physics of hydrogen isotopologues and has been recently applied to the metrology of the Q(1) 1-0 line of $H_2$.

**Introduction**
Over almost three decades, from the 70s to the 90s, Coherent Raman Scattering[1] (CRS) spectroscopy has been the approach of election for intensive spectroscopic studies of molecular hydrogen and its isotopologues, namely $D_2$ and HD. Several reasons underpinned these studies. The large Raman cross section[2] of molecular hydrogen and its sparse Raman spectrum made it among the best gas-phase candidates for the realization of Stimulated Raman amplifiers to shift the wavelength of pulsed lasers[3], [4]. In combustion[5] and plasma[6] diagnostic, $H_2$ was introduced as a probe for temperature determinations via Coherent Anti-Stokes Raman Scattering[1] (CARS), thanks to its largely separated Raman lines and to the chance to extract the temperature from their relative intensities. As the determination of intensities implied the knowledge of Raman lineshapes, calculations were carried out on the effects of collisions on the vibrational spectra[7] of $H_2$ and its main isotopologues. The calculated spectra were comparatively analyzed against experimental data mostly acquired by Stimulated Raman Scattering[1] (SRS), which offers the advantage of not distorting spectral lineshapes[8] with respect to CARS. Collisional pressure shift and broadening coefficients[9]–[13] and their temperature dependencies[14]–[16] were measured for all major isotopologues and also related to the angular and radial dependence of the molecular interaction potential[17]. To discriminate among different line-broadening mechanisms, such as those arising from elastic and inelastic collisions, CRS measurements were performed on both rotational[18], [19] and vibrational bands, because of their different collisional physics, as well as on depolarized[20], [21] (or anisotropic) and polarized (isotropic) components of Q transitions. On another front, the amenability of $H_2$ to quantum calculations of its energy levels[22] led to the first comparisons between experimental and calculated energies and to the development of more and more refined lineshape models to extrapolate zero-pressure line centers from collision-perturbed lineshapes[23]. At the turn of the millennium the number of $H_2$ studies started to decrease together with the use of Coherent Raman spectrometers.

A renewed interest for $H_2$ rovibrational spectroscopy was triggered in 2011 thanks to an accurate list of transition frequencies obtained by Komasa and Pachucki[24] from ab-initio quantum-electrodynamic calculations. The importance of comparing this database with accurate laboratory measurements was early recognized by the group of Wim Ubachs (who had shown in 2008 [25] the potential for fundamental physics of Lyman and Werner bands of $H_2$) as a testbed for new physics, such as fifth forces[26], extra-dimensions[27] and physics beyond the Standard Model[28]. In 2019, a new refined line list appeared with relativistic corrections up to $\alpha^5 m$ bringing the accuracy of theoretical calculations to the sub-MHz level, which is the current benchmark[29]. The effect of these papers was to revitalize experimental investigations. These could benefit from the unprecedented advantage of an absolute frequency scale given by the newly invented frequency combs. A first milestone was obtained by Resonantly-Enhanced Multi-Photon Ionization[30] (REMPI): in this approach vibrationally excited molecules are selectively ionized through resonant multi-photon absorption of a pulsed ultraviolet laser and then selectively detected from the mass of the generated ion. The combination of comb calibrated laser frequencies and of a molecular beam suppressing collisional and Doppler broadening enabled the Q(J) transition frequencies of the fundamental vibrational band of $H_2$, $D_2$ and HD to be measured with an 8 MHz accuracy, almost an order of magnitude better than previous measurements performed by Fourier Transform Spectroscopy[31] and CRS[10]. Recently, this benchmark for REMPI was substantially improved for the S(0) fundamental rovibrational line of $D_2$, down to 17 kHz[32], thanks to a more efficient vibrational excitation of the molecules and to a better control of Type B errors induced by the fine structure of the molecule. In parallel, taking advantage of the enormous progress in the quality of mirrors, modulators and lasers in the telecom spectral range, a number of cavity-enhanced absorption spectrometers (CEAS) were developed to address the overtone lines of $H_2$ and its isotopologues in a Doppler broadening regime. The very small absorption cross-section of their quadrupole transitions is compensated in CEAS by a strongly increased effective absorption path length[33]. The first measurements calibrated against a simple wavemeter were successfully compared with the first theoretical line list within 20 MHz[34]–[36]. The addition of comb calibration resulted in uncertainty budgets below 1 MHz for several quadrupole lines[37]–[42], primarily of $D_2$ whose 2-0 band falls in the highly accessible telecom range. The CEAS benchmark is on the S(2) line of $D_2$, with an uncertainty of about 170 kHz[42]. In cavity measurements the lineshape model adopted for the fitting of experimental spectra was soon recognized as a limiting factor for the uncertainty budget, due to the nontrivial impact of velocity changing collisions and speed-dependent effects in a colliding environment of $H_2$ molecules[43]. The experimental progresses in the sensitivity of CEAS setups has been thus accompanied by a strong effort to improve the accuracy of lineshape models, with the testing of speed-dependent billiard ball profiles[44], the implementation of a $\beta$-corrected[45] Hartmann Tran Profile ($\beta$HTP)[46], the integration in $\beta$HTP of collisional parameters obtained through *ab initio* calculations: this enables to reduce the number of fitting parameters[47] and to assign physically meaningful values to collisional parameters showing strong correlation in the fitting. Another benefit of fixing some collisional parameters is to retrieve meaningful confidence intervals for the parameters fitted[48]. On another closely related front, namely the heteronuclear HD isotopologue that exhibits weak dipole-allowed transitions, several measurements recently attained final accuracies in the 10-150 kHz range: this happened for the fundamental R(0) line studied by REMPI on a molecular beam[49], for the R(1) 2-0 line investigated at different pressures and temperatures by both sub-Doppler[50], [51] and Doppler broadening spectroscopy[52], [53], for the recently addressed R(1), R(3), P(3)[54], R(0)[52] 2-0 lines, the latter observed at cryogenic temperatures and Pascal-level pressures in an optical cavity. For the largely studied R(1) 2-0 line, the agreement within 200 kHz of 4 completely different determinations is highly significant, also considering the difficulty to fit and interpret the dispersive lineshape of Lamb dips unexpectedly encountered in sub-Doppler measurements[55]–[57].

In the past decade, apart from REMPI measurements, fundamental quadrupole transitions have almost never been addressed experimentally, mainly due to the difficulty to achieve high-sensitivity CEAS in the mid-infrared, where the quality of lasers, mirrors, modulators and detectors is poorer, and the costs are higher. The CRS approach was chosen for the metrology of Tritium-bearing molecules[58] ($T_2$, DT, HT) and the experimental validation of calculated broadening and shift coefficients for rotational lines of $D_2$[48] and HD[59]. However, in both cases the setups were reminiscent of those developed before year 2000 for Raman spectroscopy over large temperature scales[15], [60], as based on nanosecond lasers that limit resolution and accuracy to 50 and 6 MHz[58], respectively. For CRS studies at higher resolution, the traditional approach firstly proposed by Owyung[61] and later improved by Rosasco[9], [62] and Forsman[63] was based on a single longitudinal mode Ar-ion laser for the pump (488.0 nm, 300 mW, 15 MHz FWHM bandwidth for a 1 s average) and on a single-mode tunable dye laser for the Stokes (592-593 nm, 200 mW, 1 MHz FWHM bandwidth for a 1 s average). The Raman signal was measured as stimulated Raman gain, which implies pump intensity modulation and synchronous detection of the Stokes intensity change. In the latest version of the spectrometer, Forsman et al.[63] managed to achieve a spectral resolution of 1 MHz by stabilization of the Ar laser to an external cavity, a frequency accuracy of 2 MHz by calibration of pump and Stokes frequencies against a pressure-tight temperature-stabilized off-axis Fabry Perot interferometer, and a signal-to-noise ratio (SNR) of 1000 in 1 s for the Q-branch of $D_2$ at a few amagat.

Our group, attracted by the versatility of SRS to address both vibrational and rotational transitions with well assessed near-infrared technology and by the chance to calibrate pump and Stokes laser frequencies against a frequency comb for maximum resolution and accuracy, devised the potential to improve the old layouts and develop a metrology-grade SRS spectrometer for quadrupole lines of homonuclear species and in general of Raman active lines. We have recently applied it to determine the transition frequency of the prototypical Q(1) line of the 1-0 band of $H_2$ at ≈ 4155 cm$^{-1}$ with a combined uncertainty of $1.0·10^{-5}$ cm$^{-1}$ (310 kHz [64]), improving by 20 times the experimental benchmark[30] and by a factor of 2 the theoretical benchmark[29]. This result comes from a frequency accuracy improved by a factor of 31 (65 kHz against 2 MHz) and by a signal-to-noise ratio increased by a factor of 8 as compared to Forsman et al.[63]. This paper provides a detailed description of the spectrometer and of its performance, by sequentially analyzing all major parts of the apparatus, namely excitation and calibration laser sources, optical beamlines, gas chamber, comb referencing of pump and Stokes lasers, detection chain, procedures for acquisition, averaging and calibration of SRS spectra.

**Comb-calibrated SRS spectrometer**

*General layout.* The spectrometer relies on a SRS process driven by two narrow-linewidth CW lasers whose frequency is calibrated against an optical frequency comb. The comb provides repeatability and absolute calibration of the detuning between pump and Stokes frequencies. The signal-to-noise ratio (SNR) is maximized by use of a multipass cell that enhances the interaction length between gas and laser fields and by implementing a detection chain that works at the shot noise limit: this limit is obtained by modulating at high frequency (several MHz) the intensity of the Stokes laser and by performing lock-in detection of the stimulated Raman loss (SRL) imparted on the pump beam. Contributions to systematic errors from variations of thermodynamic parameters of the sample are minimized by active stabilization of both temperature and pressure of the gas, while spectral distortions induced by power changes of the Stokes laser while it is scanned across the Raman transition are quenched by an active stabilization of the Stokes power. Finally, as misalignment between pump and Stokes lasers was found to be responsible for systematic shifts of the measured line center frequency, we introduced a system for active stabilization of the overlap between pump and Stokes beams in the multipass cell. The layout of the spectrometer is depicted in Fig. 1. It is composed of several parts discussed in detail in the following, namely laser sources, optical beam lines, multipass cell, comb referencing, detection chain, SRL spectra acquisition, spectra averaging and calibration and active beam alignment.

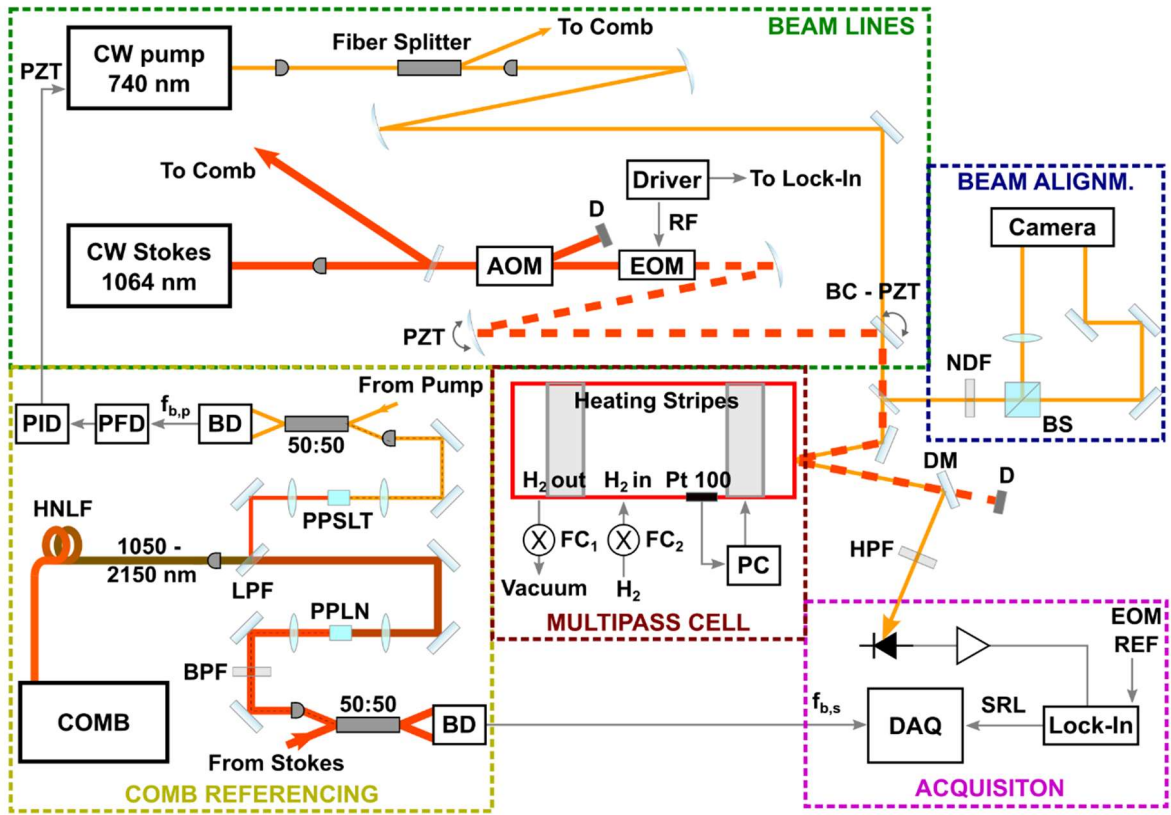

**FIGURE 1. Optical layout of SRS the spectrometer.** Colored boxes encase the parts of the setup described in detail in the sections of the main text. AOM: acousto-optic modulator; D: (beam) dump; EOM: electro-optical modulator; RF: radiofrequency signal; PZT: piezoelectric actuator; BC: beam combiner; DM: dichroic mirror; HPF: high-pass filter; BS: beam splitter; NDF: neutral density filter; SRL: stimulated Raman loss signal; DAQ: digital acquisition board; BD: balanced detector; LPF: low-pass filter; BPF: band-pass filter; PPLN: periodically-poled lithium niobate crystal; PPSLT: periodically-poled stoichiometric lithium tantalate crystal; PFD: phase-frequency detector; PID: proportional-integral-derivative controller.

*Laser sources.* The spectrometer makes use of three lasers sources, namely: i) an external-cavity (EC) diode laser (Toptica DL pro) as a pump beam of the SRS process, with tunability from 710 to 740 nm and power up to 40 mW; ii) an amplified distributed-feedback (DFB) Ytterbium fiber laser at 1064 nm (Koheras Boostik HP) as Stokes beam, featuring single-mode operation, piezo-electric frequency tuning over 15 GHz and optical power up to 15 W; iii) an Erbium:fiber mode-locked oscillator at 100 MHz (Menlo C-Fiber) followed by a home-made optical amplifier and supercontinuum stage for frequency comb calibration of pump and Stokes laser frequencies. This laser configuration is favorable to operate the spectrometer in the so-called inverse-Raman scattering regime, in which an intense Stokes laser is modulated and the SRL signal on the pump beam is detected. The choice of an EC diode laser for the pump beam brings several advantages: i) a high SNR on the measured SRL, thanks to a shot-noise-limited detection (see section "Detection chain") and to the high quantum efficiency of silicon detectors around 700 nm; ii) the coverage of several Raman transitions, specifically all fundamental lines of the Q branch of $H_2$ around 4155 $cm^{-1}$ (739 nm) and the S(0) line at 4497 $cm^{-1}$ (720 nm), thanks to the broad wavelength tuning range; iii) a robust frequency locking to the comb thanks to the large dynamic range (few gigahertz) and high-bandwidth (kHz level) of the piezo-actuated frequency tuning port. During the spectral measurements, we keep the pump locked to the nearest comb tooth and exploit the piezo-modulation port of the Stokes laser to modify the detuning between the two cw lasers. This can be done over a 15 GHz range that fully covers Raman spectra at both low and high pressures.

***Optical beamlines.*** The pump beamline starts with a single-mode optical fiber to spatially filter the beam and remove the laser astigmatism. The beam circularity is crucial to match it to the nearly confocal multipass cell used as a gas chamber and to the co-propagating Stokes beam. At the fiber output the collimated pump beam passes through a telescope that shapes the beam to an optimized beam-waist radius of 220 μm in the middle of the cell. A fraction of the pump beam is split out from the initial fiber patch and sent to the comb calibration part of the setup. The Stokes laser does not require any spatial filtering because of the intrinsically high spatial quality guaranteed by the fiber format. A beam sampler splits a small fraction of it towards the comb calibration unit, while the major fraction sequentially crosses an acousto-optic modulator for power stabilization, an electro-optic intensity modulator (R7v-10R3-YAG from Qubig) used for the synchronous detection of the SRL signal, and a telescope to optimize the injection into the multipass cell (with a beam waist of 265 μm at its center). Reflecting optics are used for all telescopes to quench parasitic etaloning effects that might alter the flatness of the spectral baseline. Stokes and pump beams are recombined before the cell by a dichroic mirror and brought to the same linear polarization state by a Glan-Thompson polarizer to avoid any distortion of the SRL response[1]. At the cell output a prism extracts the pump beam and redirects it to an amplified home-built silicon photodiode for SRL detection. A notch filter at 1064 nm protects the photodiode from the Stokes stray light, avoiding any undesired signal background.

***Gas chamber.*** The gas chamber is constituted by a multipass cell with a geometrical length of 42 cm that provides, after 70 bounces, an effective interaction length $L = 30$ m. This enhances the SRL signal since the SRS processes is perfectly phase matched along the whole interaction length[1]. The nearly confocal cell configuration (Herriot configuration) sets for the recirculating beam injected under optimal conditions a spot radius changing from a minimum value $w_0 = \sqrt{\lambda L/2\pi}$ at the cell center to $\sqrt{2}w_0$ at the cell mirrors[65]. The cell can contain gas in a pressure range from $10^{-3}$ to 5 bar and is equipped with broadband dielectric mirrors that guarantee a total transmission around 50% from 700 to 1100 nm. In typical conditions the injected Stokes power is 3 W while the pump power at the SRL detector, which is relevant to compute the shot noise, is 350 μW. Pressure and temperature of the gas inside the cell are actively stabilized to ensure stable thermodynamic conditions and to allow an efficient averaging of multiple spectra acquired over long measurement times. A temperature uniformity better than 100 mK results from the thermal conductivity of the steel that the cell is made of, combined with a surrounding box made with thick Styrofoam and equipped with internal air circulation by two fans. The temperature is measured by a calibrated Pt100 probe and a 6 ½ digit multimeter with an overall accuracy of 50 mK. A LabView-based PID servo maintains a temperature stability < 30 mK by regulating the current passing through stripe heaters glued onto the cell. To maintain a constant pressure during the measurements and compensate for small leaks of the cell, a constant flow of about $10^{-2}$ L/min is established in the cell using two flow controllers, one at the gas inlet and another one at the cell output upstream the vacuum pump. The pressure inside the cell is measured via a calibrated pressure sensor with relative uncertainty better than $10^{-3}$. Through a software PID control loop, the output flow is regulated to maintain a constant pressure inside the cell within 0.1 mbar.

***Comb referencing.*** The comb referencing of pump and Stokes lasers is obtained by generating their respective beat notes (BNs) with frequency-doubled spectral portions of an octave-spanning comb supercontinuum: specifically, the Stokes laser is made to beat with the second harmonic of the 2128 nm part of the continuum while the pump laser with the second harmonic of the 1480 nm part of the continuum. Second harmonic generation (SHG) takes place in a periodically-poled lithium niobate crystal and in a periodically-poled lithium tantalate optical waveguide for Stokes and pump, respectively: the waveguiding medium is used because of its larger conversion efficiency to compensate for the small power spectral density of the comb around 1480 nm. This referencing scheme requires neither the knowledge nor the stabilization of the carrier-envelope frequency ($f_{ceo}$) of the comb to achieve an absolute measurement of the frequency detuning between pump and Stokes lasers: this is because (see section "Spectra calibration and averaging") their beat-note signals are affected by the same $2f_{ceo}$ term arising from SHG, which cancels out in the subtraction of the two frequencies. The repetition frequency $f_{rep}$ is thus the only comb parameter stabilized, against a GPS-disciplined Rb oscillator that acts as a master clock to calibrate also the beat notes. The stability of this clock is at a level $10^{-11}$ at 1 s, thus far in excess of the $10^{-9}$ frequency uncertainty limit of the spectrometer set by the beam pointing instability of the lasers (see section "Active alignment of the laser beams"). The instrumental broadening of the spectrometer is almost negligible thanks a short-term linewidth < 100 kHz for both pump and Stokes lasers, as it is shown in Fig. 2 by the respective beat notes with the comb. The spectral resolution of the SRS spectrometer exceeds by a factor of 10 that of the best previous realizations based on Argon and dye lasers.

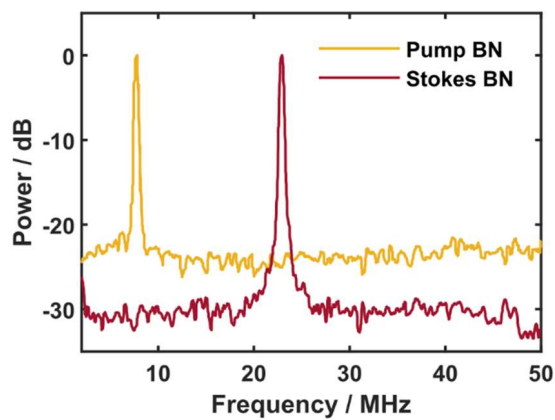

**FIGURE 2. Beat notes electrical spectra.** Spectra are normalized to bring both peaks at the 0 dB level for easier visual comparison.

*Detection chain.* The Stimulated Raman Loss signal corresponds to a Stokes-induced intensity change of the pump beam. In our experimental conditions it remains smaller than $10^{-3}$, even at high pressure on the intense fundamental Q(1) line of $H_2$. To measure it at high signal-to-noise ratio (SNR) we recurred, as it is typical in any SRS measurement, to a synchronous detection approach in which the Stokes laser beam is intensity modulated at high frequency (9.7 MHz, in our case) and a lock-in amplifier extracts the amplitude of the modulation signal transferred to the pump beam. We evaluated the SNR of the detection chain by exploiting a built-in functionality of the adopted lock-in amplifier (H2FLI from Zurich Instruments) to measure, in a 1 Hz bandwidth, the relative intensity noise spectrum of the pump laser on the SRL photodiode, with no Stokes irradiation on the sample and thus no Raman loss. Figure 3a shows this spectrum as the ratio between the average root-mean-squared noise voltage measured by the lock-in amplifier and the DC voltage at the detector, thus in units that can be directly compared to the SRL. Three out of the five spectra shown in the figure are experimental and refer to the lock-in noise background (blue line), the lock-in plus detector noise background (i.e. with pump off, red line), and the noise under pump irradiation (yellow line), thus with the addition of laser intensity and shot noise. While the noise from the lock-in is negligible, the noise from the detector is slightly smaller than the theoretical shot noise floor of $4.1 \cdot 10^{-8}$ Hz$^{-0.5}$ (purple line) calculated for a pump optical power of 350 μW, which we use in the experiments. Very importantly, at the Fourier frequency of 9.7 MHz used in our experiments, the intensity noise of the EC laser is above the shot noise level by only a factor of 1.1, indicating a very modest impact from the laser intensity noise. This can be better appreciated by calculating the quadrature sum of detector noise plus theoretical shot noise floor (green line), which closely approaches the total noise measured (at 9.7 MHz). We can thus conclude that the laser is shot noise-limited at high Fourier frequencies and that the detection chain works close to the quantum limit. An even higher sensitivity could be obtained with a less noisy detector and/or by increasing the pump laser power to reduce the shot noise floor (with an inverse square root behavior). If we compare the total noise measured of $6.1 \cdot 10^{-8}$ Hz$^{-0.5}$ with the measured rms SRL signal peak of $7.6 \cdot 10^{-4}$, which holds for the Q(1) of $H_2$ at a pressure of 1 atm, we obtain a highly favorable SNR of 8000 on a single spectral point over a 1 s measurement time (approximately corresponding to the 1 Hz bandwidth). This is by factor of 8 better than any previous SRS spectrometer. Figure 3b reports typical spectra acquired at different pressures on the Q(1) fundamental rovibrational line of $H_2$. The measurement time varies from a minimum of 5 min at high pressure to a maximum of 1 h at low pressure. The figure legend reports the SNR of the measured spectra when the spacing between spectral points is 1 MHz: even at lower pressures, where the SNR is reduced by the lower gas density and the larger profile (dominated by Doppler broadening rather than by Dicke narrowing), the SNR is consistent with a statistical uncertainty on the line center lower than 0.5 MHz, thus suitable for optical metrology.

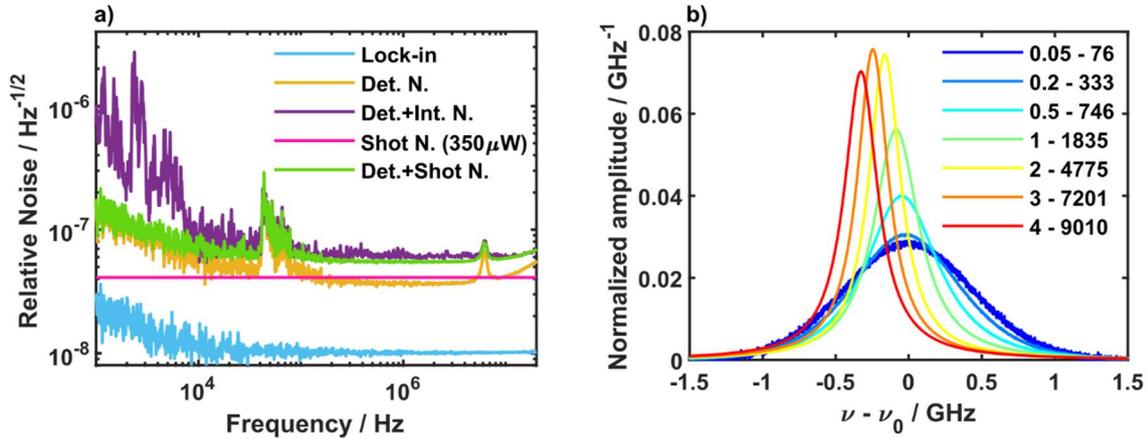

**FIGURE 3. Noise performance.** a) Comparison of the detection noise with the shot noise level. From top to bottom in the legend: lock-in noise floor, detector noise (Det. N), detector + laser intensity noise(Int. N), shot noise calculated for the pump power impinging on the detector (350 µW), detector + shot noise added in quadrature. b) Normalized SRL spectra of the Q(1) 1-0 line of $H_2$ measured at different pressures. The first number in the legend represents the measurement pressure in bar, the second one is the SNR of the spectrum.

*Spectra acquisition.* Spectral measurements are preceded by a coarse adjustment of the pump wavelength by means of an optical spectrum analyzer to match the pump-Stokes frequency detuning to the target vibrational frequency. The pump laser frequency is then offset-locked to the nearest comb mode, 10 MHz apart. The spectra are acquired by measuring the SRL signal with a 1 µs lock-in integration time while scanning the Stokes frequency over about 12 GHz around the center of the transition. Depending on the gas pressure and thus on the magnitude of the SRL signal, the frequency scans are repeated at rates of 0.1 Hz or 1 Hz, corresponding to frequency tuning speeds of ~2.4 and ~24 kHz/µs, respectively. For the calibration of the comb frequency axis the comb-Stokes beat note is synchronously digitized with the SRL signal at a fast rate using a 100 MSamples s$^{-1}$ 14 bit National Instrument PXIe-7961 board. Its onboard FPGA processor (NI-5781) allows the beat note frequency to be calculated in real time every 10 µs by Fast Fourier Transform processing of data segments composed of 1024 samples. This is equivalent to having SRL points spectrally separated by 24 or 240 kHz depending on the adopted scan rate. The absolute frequency is reconstructed in post-processing by unwrapping the measured beat note frequency with the procedure described in detail in the next section. The total acquisition time of an SRL spectrum for a given line and under given thermodynamic conditions varies from 5 to 30 min depending on the pressure. Longer times can be implemented to further enhance the SNR thanks to the robustness of the comb calibrated frequency measurement.

It is worth emphasizing that without a proper power stabilization of both pump and Stokes lasers the shot-noise limit above discussed is far from being the ultimate limitation to the quality of an SRL spectrum. Spectral distortions may easily occur because in an SRS process the measured pump intensity change is linearly proportional to pump and Stokes powers that can fluctuate appreciably over time during spectral acquisitions (even if these are performed at a fast rate). Apart from the normal power drift of any laser, Stokes power fluctuations occur due to the piezo-actuated frequency scans and to the presence of parasitic etalons. Figure 4 quantifies this issue by reporting the time-dependent power of the Stokes laser when its frequency is kept fixed (panel a, blue trace), when it is varied at 1 Hz in a typical scan (panel b, blue trace), and when a proper active power stabilization is switched on (red trace in both panels): the fluctuations are of the order of $10^{-2}$ without stabilization and $10^{-4}$ with active stabilization. The power stabilization is thus mandatory to preserve the shot-noise limit in SRL spectra and to keep distortions of the spectral lineshape below the noise level. It is implemented via an acousto-optic modulator (AOM) whose diffraction efficiency is controlled by a PID servo to stabilize the power of the $0^{th}$ diffraction order on a monitoring detector.

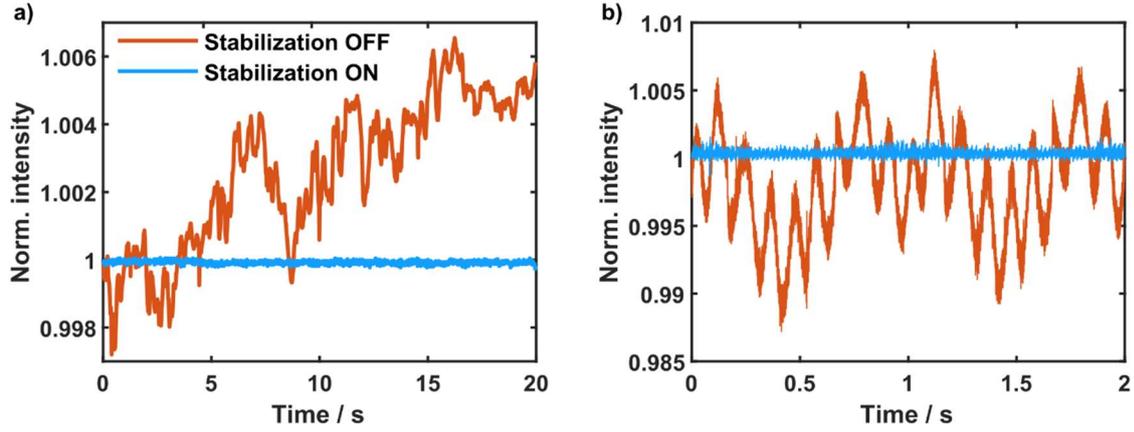

**FIGURE 4. Active stabilization of laser power.** Power fluctuations of the Stokes beam relative to the power at t = 0 s when the active stabilization is inactive (red trace) or active (blue trace), in the case of fixed Stokes laser frequency (a) and scanning Stokes frequency (b) over about 10 GHz at a rate of 1 Hz.

***Spectra calibration and averaging.*** The absolute calibration of spectra is performed by assigning to each spectral point the corresponding value of the Raman detuning $\Omega = \nu_p - \nu_s = n_p f_{\text{rep}} \pm f_{\text{bn,p}} - (n_s f_{\text{rep}} \pm f_{\text{bn,s}})$, where $\nu_p$ and $\nu_s$ are the optical frequencies of pump and Stokes lasers, respectively, which can be written as an integer number times the comb repetition frequency ($n_x f_{\text{rep}}$) plus or minus the beat note $f_{bn,x}$ between the laser and the frequency comb. The carrier-envelope offset frequency of the comb has been omitted as it equally affects the comb teeth used for pump and Stokes referencing and thus cancels out in the subtraction $\nu_p - \nu_s$. The frequency detuning can be expressed in compact form as $\Omega = \Delta n\, f_{\text{rep}} \pm f_{\text{bn,p}} \pm f_{\text{bn,s}}$, where $\Delta n = n_p - n_s$. The sign of $f_{\text{bn,s}}$ is easily assessed by the known direction of the scan, the sign of $f_{bn,p}$ is determined by the sign of the lock, while $\Delta n$ can be determined minimizing the discrepancy between the measured and theoretical transition frequency, which is known with an uncertainty much lower than $f_{rep}$.

The acquisition board acquires synchronously the SRL signal together with the value of $f_{bn,s}$ in the range $0 \div f_{rep}/2$, which corresponds to 50 MHz in our case. The result is a beat note frequency which follows a sawtooth pattern as shown in Fig. 5a. The measured $f_{\text{bn,s}}$ is then unwrapped to retrieve a monotonically increasing frequency, representing the relative frequency of the Stokes laser with respect to a comb tooth (Fig. 5b). When $f_{\text{bn,s}}$ is close to DC or to $f_{\text{rep}}/2$, the beat note power is too low for a correct determination of its frequency, due to AC coupling and low-pass filtering below 50 MHz to avoid aliasing. To compute the beat-note in these blank regions we perform a quadratic interpolation of $f_{\text{bn,s}}$ in an interval of 50 MHz around the blank region. From the standard deviation of the residuals of these repeated quadratic fits used for the interpolations of $f_{\text{bn,s}}$ over time we estimate a precision of 80 kHz for the frequency calibration of each single spectral point.

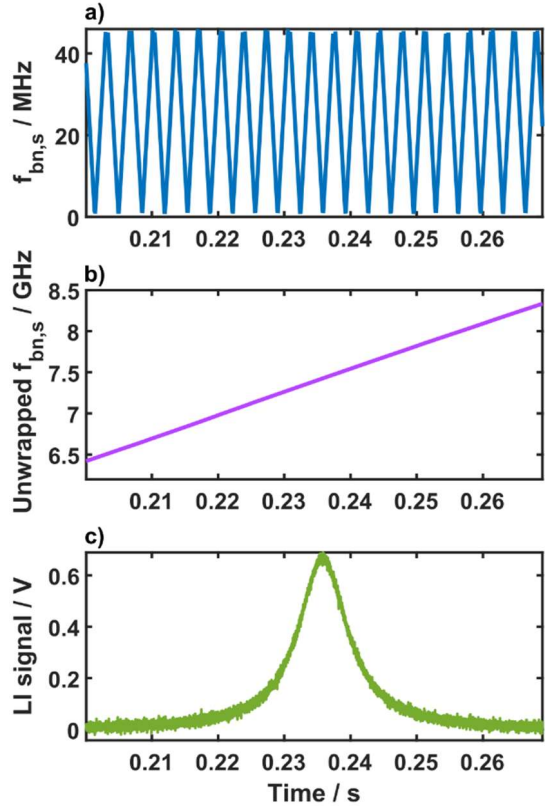

**FIGURE 5. Stokes frequency calibration.** a) Time-trace of the beat note frequency $f_{bn,s}$ between Stokes laser and comb during a spectral acquisition, as measured by the acquisition card. b) Unwrapped $f_{bn,s}$ used to calibrate the spectral frequency axis. c) SRL signal synchronously acquired by the card.

Once single spectra are calibrated in frequency, they can be combined to produce an average spectrum. As their points are not acquired on a regular, evenly spaced frequency grid, we perform a binning of spectral points by dividing the frequency axis in 1-MHz wide bins and by averaging all the points falling inside the same bin.

***Active alignment of the laser beams.*** We experimentally found that a small angular tilt between pump and Stokes beams translated into a shift of their actual frequency detuning. In our setup, manual alignment could guarantee an accuracy up to 300 μrad, which corresponds to frequency fluctuations of more than 1 MHz on repeated measurements.

To reduce this source of uncertainty, we implemented an active alignment of the Stokes beam onto the pump. The system is illustrated in Fig. 6a: the superimposed pump and Stokes beams are sampled right after the dichroic beam combiner, then further split into two replicas impinging onto different regions of a CMOS colour camera. The first replica passes through a lens that images the plane of the beam combiner onto the sensor, while the second replica is made to propagate a total distance equal to that between the beam combiner and the centre of the cell before hitting the camera. We may refer to these planes as the near and far field (NF and FF, respectively).

The RGB colour filters of the camera sensor have different responses to the wavelengths of the pump and Stokes beams: the pump is maximally transmitted through the red channel, while the Stokes is equally transmitted through all channels. We can model the camera detection of the two beams through a matrix that maps the local intensity of the pump and Stokes beams, $I_p$ and $I_S$, respectively, onto the RGB signals of the corresponding pixel:

$$\begin{pmatrix} R \\ G \\ B \end{pmatrix} = M \begin{pmatrix} I_p \\ I_S \\ 0 \end{pmatrix}.$$

The matrix $M$ can be determined column by column by sending one beam at a time on the camera sensor and measuring the response of each colour channel. The above equation can then be inverted to retrieve the local intensity of the superimposed pump and Stokes beams:

$$\begin{pmatrix} I_p \\ I_S \\ U \end{pmatrix} = M^{-1} \begin{pmatrix} R \\ G \\ B \end{pmatrix},$$

where $U$ represents an irrelevant output. Vertical and horizontal intensity profiles of both beams are then fitted with a Gaussian function that retrieves their position in the NF and FF. Thanks to four PID controllers implemented in LabView that act on piezoelectric actuators placed on two tip-tilt mirror mounts controlling transverse position and tilt of the Stokes beam, an active alignment of the Stokes onto the pump beam is eventually performed both in the NF and FF. Fig. 6b shows the effect of manual and automatic beam alignment on the retrieved centre frequency. Automatic beam alignment results in line centre fluctuations upon repeated measurements reduced to 65 kHz rms, at the level dictated by the statistical noise, as compared to fluctuations of more than 1 MHz with manual alignment.

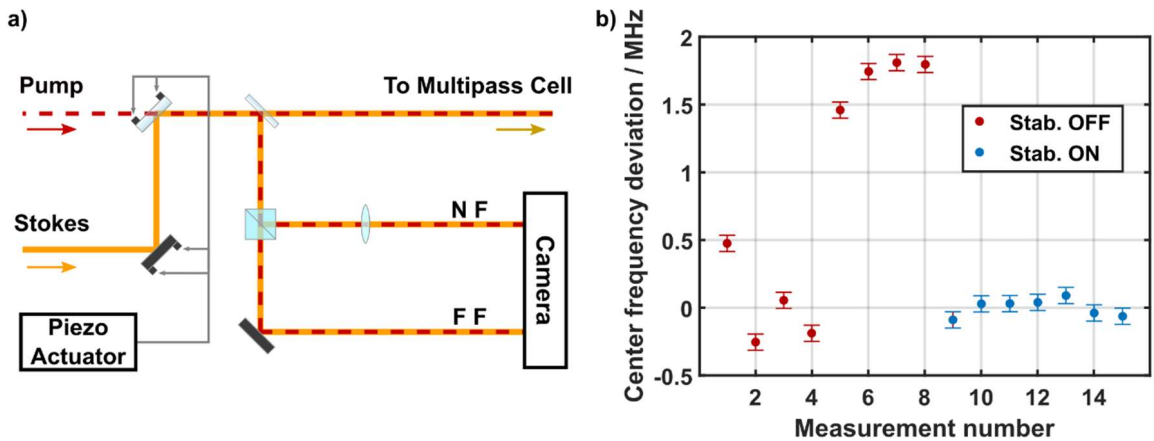

**FIGURE 6. Active beam alignment.** a) Schematic diagram of the active beam alignment system; NF: near field; FF: far field. b) Fluctuations of the line center frequency over different measurements with the gas kept at constant thermodynamic conditions and the Stokes beam misaligned and realigned onto the pump beam before each measurement. Red dots (measurements 1-8) represent measurements where the alignment is manual, while blue dots (9-15) represent measurements under active beam alignment. Each dot corresponds to an averaged spectrum acquired over 5 min, while error bars are the standard deviation of the mean. The vertical scale is relative to the mean line center frequency retrieved with active stabilization.

**Conclusions**

We have presented and discussed in detail a Stimulated Raman Loss spectrometer for optical metrology of Raman-active transitions and in particular of weak quadrupole transitions that are relevance for fundamental physics. It makes use for the first time of an optical frequency comb to achieve a repeatable and accurate frequency axis calibration. The nonlinearity of the approach makes it suitable to probe with near-infrared lasers fundamental transitions that could be hardly addressed with similar signal-to-noise ratios by absorption spectrometers operating in the mid-infrared. The choice of an SRS regime adds a very high spectral fidelity that makes it of interest also for highly accurate measurements of molecular lineshapes and thus for collisional studies. The metrological attitude of the spectrometer was recently demonstrated by the measurement of the transition frequency of the Q(1) 1-0 line of H2 at 4155 cm$^{-1}$ with an accuracy of few parts-per-billion (1.0·10$^{-5}$ cm$^{-1}$), corresponding to an improvement by a factor of 10 and 2, respectively, of the current experimental and theoretical benchmarks. With minor technical changes, namely the replacement of the pump laser, it is susceptible to address a two decades-spanning frequency range, from 50 to 5000 cm$^{-1}$, that covers all fundamental rovibrational bands as well as purely rotational bands.

As it is a nonlinear spectrometer where the signal scales linearly with the laser intensity and the interaction length, one may easily anticipate a substantial boost of the performance by replacement of the multipass cell with a hollow-core photonic crystal fiber (HC-PCF). In a multipass cell an effective interaction length of few tens of meters comes along with a spot size diameter of about 300 µm. An HC-PCF, which can provide the same interaction length with a 10 times smaller spot-size, is able to increase the intensity, and thus the signal-to-noise ratio by a factor of 100. This will be highly suitable to perform measurements at very low pressures [52] or under dilution with suitable collisional perturbers (e.g. He, Ar)[66] to further reduce the uncertainty budget on the final transition frequencies. The fiber environment is also of extreme interest to enter collisional regimes dominated by wall collisions that can significantly simplify the regression to zero pressure of the measured line center frequencies. The upgrade of the spectrometer in this fiber perspective is currently in progress.

REFERENCES

[1] G. L. Eesley, "Coherent raman spectroscopy," *J Quant Spectrosc Radiat Transf*, vol. 22, no. 6, pp. 507–576, Dec. 1979, doi: 10.1016/0022-4073(79)90045-1.

[2] A. D. Devir, "The determination of absolute Raman cross sections by indirect measurement of stimulated Raman gain," *J Appl Phys*, vol. 49, no. 6, pp. 3110–3113, Jun. 1978, doi: 10.1063/1.325301.

[3] D. Hanna, D. Pointer, and D. Pratt, "Stimulated Raman scattering of picosecond light pulses in hydrogen, deuterium, and methane," *IEEE J Quantum Electron*, vol. 22, no. 2, pp. 332–336, Feb. 1986, doi: 10.1109/JQE.1986.1072945.

[4] A. Kazzaz, S. Ruschin, I. Shoshan, and G. Ravnitsky, "Stimulated Raman scattering in methane-experimental optimization and numerical model," *IEEE J Quantum Electron*, vol. 30, no. 12, pp. 3017–3024, 1994, doi: 10.1109/3.362703.

[5] W. Stricker, M. Woyde, R. Lückerath, and V. Bergmann, "Temperature Measurements in High Pressure Combustion," *Berichte der Bunsengesellschaft für physikalische Chemie*, vol. 97, no. 12, pp. 1608–1618, Dec. 1993, doi: 10.1002/bbpc.19930971217.

[6] M. Pealat, J. P. E. Taran, J. Taillet, M. Bacal, and A. M. Bruneteau, "Measurement of vibrational populations in low-pressure hydrogen plasma by coherent anti-Stokes Raman scattering," *J Appl Phys*, vol. 52, no. 4, pp. 2687–2691, Apr. 1981, doi: 10.1063/1.329075.

[7] D. Robert, J. Bonamy, J. P. Sala, G. Levi, and F. Marsault-Herail, "Temperature dependence of the vibrational phase relaxation in gases: Application to H$_2$-rare gas mixtures," *Chem Phys*, vol. 99, no. 2, pp. 303–315, Oct. 1985, doi: 10.1016/0301-0104(85)80127-0.

[8] D. Polli, V. Kumar, C. M. Valensise, M. Marangoni, and G. Cerullo, "Broadband Coherent Raman Scattering Microscopy," *Laser Photon Rev*, vol. 12, no. 9, p. 1800020, Sep. 2018, doi: 10.1002/lpor.201800020.

[9]  G. J. Rosasco, A. D. May, W. S. Hurst, L. B. Petway, and K. C. Smyth, "Broadening and shifting of the Raman $Q$ branch of HD," *J Chem Phys*, vol. 90, no. 4, pp. 2115–2124, Feb. 1989, doi: 10.1063/1.456005.

[10] L. A. Rahn and G. J. Rosasco, "Measurement of the density shift of the $H_2$ $Q$ (0–5) transitions from 295 to 1000 K," *Phys Rev A (Coll Park)*, vol. 41, no. 7, pp. 3698–3706, Apr. 1990, doi: 10.1103/PhysRevA.41.3698.

[11] L. A. Rahn, R. L. Farrow, and G. J. Rosasco, "Measurement of the self-broadening of the $H_2$ $Q$ (0–5) Raman transitions from 295 to 1000 K," *Phys Rev A (Coll Park)*, vol. 43, no. 11, pp. 6075–6088, Jun. 1991, doi: 10.1103/PhysRevA.43.6075.

[12] P. M. Sinclair, P. Duggan, M. le Flohic, J. W. Forsman, J. R. Drummond, and A. D. May, "Broadening and shifting of the Raman $Q$ branch in pure $D_2$ and $D_2$–He mixtures, I: experimental results and comparison with theory," *Can J Phys*, vol. 72, no. 11–12, pp. 885–890, Nov. 1994, doi: 10.1139/p94-116.

[13] P. M. Sinclair, P. Duggan, J. W. Forsman, J. R. Drummond, and A. D. May, "Broadening and shifting of the Raman $Q$ branch in $D_2$ and $D_2$–He mixtures, II: line shapes, fitting routines, and effects nonlinear in density," *Can J Phys*, vol. 72, no. 11–12, pp. 891–896, Nov. 1994, doi: 10.1139/p94-117.

[14] W. K. Bischel and M. J. Dyer, "Temperature dependence of the Raman linewidth and line shift for the Q(1) and Q(0) transitions in normal and para-$H_2$," *Phys Rev A (Coll Park)*, vol. 33, no. 5, pp. 3113–3123, May 1986, doi: 10.1103/PhysRevA.33.3113.

[15] J. Ph. Berger, R. Saint-Loup, H. Berger, J. Bonamy, and D. Robert, "Measurement of vibrational line profiles in $H_2$–rare-gas mixtures: Determination of the speed dependence of the line shift," *Phys Rev A (Coll Park)*, vol. 49, no. 5, pp. 3396–3406, May 1994, doi: 10.1103/PhysRevA.49.3396.

[16] P. M. Sinclair *et al.*, "Collisional broadening and shifting parameters of the Raman $Q$ branch of $H_2$ perturbed by $N_2$ determined from speed-dependent line profiles at high temperatures," *Phys Rev A (Coll Park)*, vol. 54, no. 1, pp. 402–409, Jul. 1996, doi: 10.1103/PhysRevA.54.402.

[17] J. D. Kelley and S. L. Bragg, "Effect of collisions on line profiles in the vibrational spectrum of molecular hydrogen," *Phys Rev A (Coll Park)*, vol. 34, no. 4, pp. 3003–3014, Oct. 1986, doi: 10.1103/PhysRevA.34.3003.

[18] X. Michaut, R. Saint-Loup, H. Berger, M. L. Dubernet, P. Joubert, and J. Bonamy, "Investigations of pure rotational transitions of $H_2$ self-perturbed and perturbed by He. I. Measurement, modeling, and quantum calculations," *J Chem Phys*, vol. 109, no. 3, pp. 951–961, Jul. 1998, doi: 10.1063/1.476638.

[19] M. P. le Flohic, P. Duggan, P. M. Sinclair, J. R. Drummond, and A. D. May, "Collisional broadening and shifting of the pure rotational Raman lines $S_0$ ($J$=0–4) of $H_2$ at room temperature," *Can J Phys*, vol. 72, no. 5–6, pp. 186–192, May 1994, doi: 10.1139/p94-029.

[20] R. L. Farrow and G. O. Sitz, "Coherent anti-Stokes Raman spectroscopy investigation of the anisotropic vibrational transitions of hydrogen," *Journal of the Optical Society of America B*, vol. 6, no. 5, p. 865, May 1989, doi: 10.1364/JOSAB.6.000865.

[21] P. M. Sinclair, P. Duggan, J. R. Drummond, and A. D. May, "Broadening and shifting of the depolarized component of the Raman $Q$ branch in $D_2$ at room temperature," *Can J Phys*, vol. 73, no. 7–8, pp. 530–536, Jul. 1995, doi: 10.1139/p95-077.

[22] C. Schwartz and R. J. le Roy, "Nonadiabatic eigenvalues and adiabatic matrix elements for all isotopes of diatomic hydrogen," *J Mol Spectrosc*, vol. 121, no. 2, pp. 420–439, Feb. 1987, doi: 10.1016/0022-2852(87)90059-2.

[23] D. A. Shapiro, R. Ciurylo, R. Jaworski, and A. D. May, "Modeling the spectral line shapes with speed-dependent broadening and Dicke narrowing," *Can J Phys*, vol. 79, no. 10, pp. 1209–1222, Oct. 2001, doi: 10.1139/p01-080.

[24] J. Komasa, K. Piszczatowski, G. Łach, M. Przybytek, B. Jeziorski, and K. Pachucki, "Quantum Electrodynamics Effects in Rovibrational Spectra of Molecular Hydrogen," *J Chem Theory Comput*, vol. 7, no. 10, pp. 3105–3115, Oct. 2011, doi: 10.1021/ct200438t.

[25] E. J. Salumbides *et al.*, "Improved Laboratory Values of the $H_2$ Lyman and Werner Lines for Constraining Time Variation of the Proton-to-Electron Mass Ratio," *Phys Rev Lett*, vol. 101, no. 22, p. 223001, Nov. 2008, doi: 10.1103/PhysRevLett.101.223001.

[26] E. J. Salumbides, J. C. J. Koelemeij, J. Komasa, K. Pachucki, K. S. E. Eikema, and W. Ubachs, "Bounds on fifth forces from precision measurements on molecules," *Physical Review D*, vol. 87, no. 11, p. 112008, Jun. 2013, doi: 10.1103/PhysRevD.87.112008.

[27] E. J. Salumbides, A. N. Schellekens, B. Gato-Rivera, and W. Ubachs, "Constraints on extra dimensions from precision molecular spectroscopy," *New J Phys*, vol. 17, no. 3, p. 033015, Mar. 2015, doi: 10.1088/1367-2630/17/3/033015.

[28] W. Ubachs, J. C. J. Koelemeij, K. S. E. Eikema, and E. J. Salumbides, "Physics beyond the Standard Model from hydrogen spectroscopy," *J Mol Spectrosc*, vol. 320, pp. 1–12, Feb. 2016, doi: 10.1016/j.jms.2015.12.003.

[29] J. Komasa, M. Puchalski, P. Czachorowski, G. Łach, and K. Pachucki, "Rovibrational energy levels of the hydrogen molecule through nonadiabatic perturbation theory," *Phys Rev A (Coll Park)*, vol. 100, no. 3, p. 032519, Sep. 2019, doi: 10.1103/PhysRevA.100.032519.

[30] M. L. Niu, E. J. Salumbides, G. D. Dickenson, K. S. E. Eikema, and W. Ubachs, "Precision spectroscopy of the $X^1\Sigma_g^+$, v=0→1(J=0–2) rovibrational splittings in $H_2$, HD and $D_2$," *J Mol Spectrosc*, vol. 300, pp. 44–54, Jun. 2014, doi: 10.1016/j.jms.2014.03.011.

[31] S. L. Bragg, W. H. Smith, and J. W. Brault, "Line positions and strengths in the $H_2$ quadrupole spectrum," *Astrophys J*, vol. 263, p. 999, Dec. 1982, doi: 10.1086/160568.

[32] A. Fast and S. A. Meek, "Precise measurement of the $D_2$ $S_1(0)$ vibrational transition frequency," *Mol Phys*, Nov. 2021, doi: 10.1080/00268976.2021.1999520.

[33] D. Romanini, I. Ventrillard, G. Méjean, J. Morville, and E. Kerstel, "Introduction to Cavity Enhanced Absorption Spectroscopy," 2014, pp. 1–60. doi: 10.1007/978-3-642-40003-2_1.

[34] S. Kassi and A. Campargue, "Electric quadrupole and dipole transitions of the first overtone band of HD by CRDS between 1.45 and 1.33 μm," *J Mol Spectrosc*, vol. 267, no. 1–2, pp. 36–42, May 2011, doi: 10.1016/j.jms.2011.02.001.

[35] S. Kassi, A. Campargue, K. Pachucki, and J. Komasa, "The absorption spectrum of $D_2$: Ultrasensitive cavity ring down spectroscopy of the (2–0) band near 1.7 μm and accurate *ab initio* line list up to 24000 cm$^{-1}$," *J Chem Phys*, vol. 136, no. 18, p. 184309, May 2012, doi: 10.1063/1.4707708.

[36] A. Campargue, S. Kassi, K. Pachucki, and J. Komasa, "The absorption spectrum of $H_2$: CRDS measurements of the (2-0) band, review of the literature data and accurate ab initio line list up to 35000 cm$^{-1}$," *Phys. Chem. Chem. Phys.*, vol. 14, no. 2, pp. 802–815, 2012, doi: 10.1039/C1CP22912E.

[37] D. Mondelain, S. Kassi, T. Sala, D. Romanini, D. Gatti, and A. Campargue, "Sub-MHz accuracy measurement of the S(2) 2–0 transition frequency of $D_2$ by Comb-Assisted Cavity Ring Down spectroscopy," *J Mol Spectrosc*, vol. 326, pp. 5–8, Aug. 2016, doi: 10.1016/j.jms.2016.02.008.

[38] E. Fasci, A. Castrillo, H. Dinesan, S. Gravina, L. Moretti, and L. Gianfrani, "Precision spectroscopy of HD at 1.38μm," *Phys Rev A (Coll Park)*, vol. 98, no. 2, p. 022516, Aug. 2018, doi: 10.1103/PhysRevA.98.022516.

[39] P. Wcisło et al., "Accurate deuterium spectroscopy for fundamental studies," *J Quant Spectrosc Radiat Transf*, vol. 213, pp. 41–51, Jul. 2018, doi: 10.1016/j.jqsrt.2018.04.011.

[40] D. Mondelain, S. Kassi, and A. Campargue, "Transition frequencies in the (2-0) band of $D_2$ with MHz accuracy," *J Quant Spectrosc Radiat Transf*, vol. 253, p. 107020, Sep. 2020, doi: 10.1016/j.jqsrt.2020.107020.

[41] S. Wójtewicz et al., "Accurate deuterium spectroscopy and comparison with *ab initio* calculations," *Phys Rev A (Coll Park)*, vol. 101, no. 5, p. 052504, May 2020, doi: 10.1103/PhysRevA.101.052504.

[42] M. Zaborowski et al., "Ultrahigh finesse cavity-enhanced spectroscopy for accurate tests of quantum electrodynamics for molecules," *Opt Lett*, vol. 45, no. 7, p. 1603, Apr. 2020, doi: 10.1364/OL.389268.

[43] P. Wcisło et al., "The implementation of non-Voigt line profiles in the HITRAN database: $H_2$ case study," *J Quant Spectrosc Radiat Transf*, vol. 177, pp. 75–91, Jul. 2016, doi: 10.1016/j.jqsrt.2016.01.024.

[44] P. Wcisło, I. E. Gordon, C.-F. Cheng, S.-M. Hu, and R. Ciuryło, "Collision-induced line-shape effects limiting the accuracy in Doppler-limited spectroscopy of $H_2$," *Phys Rev A (Coll Park)*, vol. 93, no. 2, p. 022501, Feb. 2016, doi: 10.1103/PhysRevA.93.022501.

[45] M. Konefał, M. Słowiński, M. Zaborowski, R. Ciuryło, D. Lisak, and P. Wcisło, "Analytical-function correction to the Hartmann–Tran profile for more reliable representation of the Dicke-narrowed molecular spectra," *J Quant Spectrosc Radiat Transf*, vol. 242, p. 106784, Feb. 2020, doi: 10.1016/j.jqsrt.2019.106784.

[46] J. Tennyson et al., "Recommended isolated-line profile for representing high-resolution spectroscopic transitions (IUPAC Technical Report)," *Pure and Applied Chemistry*, vol. 86, no. 12, pp. 1931–1943, Dec. 2014, doi: 10.1515/pac-2014-0208.

[47] M. Słowiński et al., "$H_2$-He collisions: *Ab initio* theory meets cavity-enhanced spectra," *Phys Rev A (Coll Park)*, vol. 101, no. 5, p. 052705, May 2020, doi: 10.1103/PhysRevA.101.052705.

[48] R. Z. Martínez, D. Bermejo, F. Thibault, and P. Wcisło, "Testing the ab initio quantum-scattering calculations for the $D_2$-He benchmark system with stimulated Raman spectroscopy," *Journal of Raman Spectroscopy*, vol. 49, no. 8, pp. 1339–1349, Aug. 2018, doi: 10.1002/jrs.5391.

[49] A. Fast and S. A. Meek, "Sub-ppb Measurement of a Fundamental Band Rovibrational Transition in HD," *Phys Rev Lett*, vol. 125, no. 2, p. 023001, Jul. 2020, doi: 10.1103/PhysRevLett.125.023001.

[50] F. M. J. Cozijn, P. Dupré, E. J. Salumbides, K. S. E. Eikema, and W. Ubachs, "Sub-Doppler Frequency Metrology in HD for Tests of Fundamental Physics," *Phys Rev Lett*, vol. 120, no. 15, p. 153002, Apr. 2018, doi: 10.1103/PhysRevLett.120.153002.

[51] L.-G. Tao *et al.*, "Toward a Determination of the Proton-Electron Mass Ratio from the Lamb-Dip Measurement of HD," *Phys Rev Lett*, vol. 120, no. 15, p. 153001, Apr. 2018, doi: 10.1103/PhysRevLett.120.153001.

[52] S. Kassi, C. Lauzin, J. Chaillot, and A. Campargue, "The (2–0) *R* (0) and *R* (1) transition frequencies of HD determined to a $10^{-10}$ relative accuracy by Doppler spectroscopy at 80 K," *Physical Chemistry Chemical Physics*, vol. 24, no. 38, pp. 23164–23172, 2022, doi: 10.1039/D2CP02151J.

[53] A. Castrillo, E. Fasci, and L. Gianfrani, "Doppler-limited precision spectroscopy of HD at 1.4μm: An improved determination of the R(1) center frequency," *Phys Rev A (Coll Park)*, vol. 103, no. 2, p. 022828, Feb. 2021, doi: 10.1103/PhysRevA.103.022828.

[54] F. M. J. Cozijn, M. L. Diouf, V. Hermann, E. J. Salumbides, M. Schlösser, and W. Ubachs, "Rotational level spacings in HD from vibrational saturation spectroscopy," *Phys Rev A (Coll Park)*, vol. 105, no. 6, p. 062823, Jun. 2022, doi: 10.1103/PhysRevA.105.062823.

[55] M. L. Diouf, F. M. J. Cozijn, B. Darquié, E. J. Salumbides, and W. Ubachs, "Lamb-dips and Lamb-peaks in the saturation spectrum of HD," *Opt Lett*, vol. 44, no. 19, p. 4733, Oct. 2019, doi: 10.1364/OL.44.004733.

[56] T.-P. Hua, Y. R. Sun, and S.-M. Hu, "Dispersion-like lineshape observed in cavity-enhanced saturation spectroscopy of HD at 1.4μm," *Opt Lett*, vol. 45, no. 17, p. 4863, Sep. 2020, doi: 10.1364/OL.401879.

[57] Y.-N. Lv *et al.*, "Fano-like Resonance due to Interference with Distant Transitions," *Phys Rev Lett*, vol. 129, no. 16, p. 163201, Oct. 2022, doi: 10.1103/PhysRevLett.129.163201.

[58] K.-F. Lai *et al.*, "Precision measurement of the fundamental vibrational frequencies of tritium-bearing hydrogen molecules: $T_2$, DT, HT," *Physical Chemistry Chemical Physics*, vol. 22, no. 16, pp. 8973–8987, 2020, doi: 10.1039/D0CP00596G.

[59] F. Thibault, R. Z. Martínez, D. Bermejo, and P. Wcisło, "Line-shape parameters for the first rotational lines of HD in He," *Mol Astrophys*, vol. 19, p. 100063, Jun. 2020, doi: 10.1016/j.molap.2020.100063.

[60] L. A. Rahn and R. E. Palmer, "Studies of nitrogen self-broadening at high temperature with inverse Raman spectroscopy," *Journal of the Optical Society of America B*, vol. 3, no. 9, p. 1164, Sep. 1986, doi: 10.1364/JOSAB.3.001164.

[61] A. Owyoung, "Coherent Raman gain spectroscopy using CW laser sources," *IEEE J Quantum Electron*, vol. 14, no. 3, pp. 192–203, Mar. 1978, doi: 10.1109/JQE.1978.1069760.

[62] G. J. Rosasco and W. S. Hurst, "Phase-modulated stimulated Raman spectroscopy," *Journal of the Optical Society of America B*, vol. 2, no. 9, p. 1485, Sep. 1985, doi: 10.1364/JOSAB.2.001485.

[63] J. W. Forsman, P. M. Sinclair, P. Duggan, J. R. Drummond, and A. D. May, "A high-resolution Raman gain spectrometer for spectral lineshape studies," *Can J Phys*, vol. 69, no. 5, pp. 558–563, May 1991, doi: 10.1139/p91-092.

[64] M. Lamperti *et al.*, "Stimulated-Raman-Scattering Metrology," Jul. 2022, doi: 10.48550/arXiv.2207.03998.

[65] J. B. McManus, P. L. Kebabian, and M. S. Zahniser, "Astigmatic mirror multipass absorption cells for long-path-length spectroscopy," *Appl Opt*, vol. 34, no. 18, p. 3336, Jun. 1995, doi: 10.1364/AO.34.003336.

[66] E. A. Serov *et al.*, "CO-Ar collisions: ab initio model matches experimental spectra at a sub percent level over a wide pressure range," *J Quant Spectrosc Radiat Transf*, vol. 272, p. 107807, Sep. 2021, doi: 10.1016/j.jqsrt.2021.107807.